\documentclass[10pt,letterpaper,twocolumn]{article} 

\usepackage{ol2_arxiv}
\usepackage[draft]{hyperref}
\usepackage{amsmath}
\usepackage{subfig}
\usepackage{tikz}
\usetikzlibrary{arrows}
\usepackage{overpic}
\def\GFT{{\bf G}}

\def\br{\mathbf{r}}

\begin{document}

\twocolumn[

\title{Finite-difference time-domain technique as an efficient tool for obtaining the regularized Green function: applications to the local field problem in quantum optics for  inhomogeneous lossy materials}

\author{C. Van Vlack$^{*}$ and S. Hughes}
\address{
Department of Physics, Queen's University, Kingston Ontario, Canada K7L 3N6\\
$^*$Corresponding authors: cvanvlack@physics.queensu.ca 
}

\begin{abstract}
 The calculation of the local density of states (LDOS) in lossy materials has long been disputed due to the divergence of the homogeneous Green function with equal space arguments. For arbitrary shaped lossy structures, such as
 those of interest in nanoplasmonics, 
 this problem is particular challenging.
 A non-divergent LDOS obtained in numerical methods like
 the finite-difference time-domain (FDTD), at first sight, appears to be wrong. 
 Here we show that FDTD is not only
 an ideal choice for obtaining the regularized LDOS, but it can address
 the local field problem for any lossy inhomogeneous material.
 We exemplify the case of a finite-size photon emitter embedded within
 and  outside a lossy metal nanoparticle, and show excellent agreement with analytical results. 
 
\end{abstract}



 ] 

\noindent
 The spontaneous emission rate is proportional to the imaginary part of the photon Green function (GF) with equal space arguments, i.e., $\Gamma \propto {\rm Im}[\GFT \left(\br,\br;\omega\right)]$, and can be decomposed into a homogeneous contribution and a scattered contribution, $\GFT = \GFT^{\rm hom} + \GFT^{\rm scatt}$. In the case of a lossless homogeneous material, then
\begin{equation}
\label{eq:Ghom}
{\rm Im}[{G}_{ii}^{\rm hom}({\bf r},{\bf r};\omega)] = k_0^3 n \mu/6\pi,
\end{equation}
while the real part diverges (here $k_0$ is the vacuum wavevector, $n$ is the refractive index, and $\mu$ is the permeability of the homogeneous space). As soon as loss is introduced (e.g., ${\rm Im}[n],{\rm Im}[\mu]  \neq 0$), then the real and imaginary parts of the GF are mixed, causing both to diverge.
This {\em problem} is well known and 
has been discussed in the context of quantum optics
 for decades~\cite{Barnet,Scheel-PRA-60-4094}.

One commonly proposed solution to this LDOS divergence has  been to consider the {local environment} around the (photon) emitter to be a {\em lossless} cavity~\cite{Scheel-PRA-60-4094,Tomas-PRA-63-053811}, thus modelling the effects of the local field as seen by the atom instead of the macroscopic field. This circumvents the issues arising due to the divergence of $\GFT({\bf r},{\bf r})$, while still including the contribution of the outer material response via $\GFT^{\rm scatt}$. For larger photon emitters such as quantum dots (QDs) and
macromolecules, the dipole model in a fictitious cavity may not be the best approach. Moreover, most of these local-field models are restricted
to ``spherical cavities,'' and are usually (but not always) applied
to homogeneous structures. For more general nanophotonic structures,
a common numerical method of choice is the finite-difference time-domain (FDTD) technique~\cite{Sullivan-FDTD}. In a lossy structure, FDTD
 obtains a  finite $\GFT({\bf r},{\bf r})$ 
for both the real and imaginary parts, and this raises the question about possible numerical problems  near and within lossy
structures. 
This question of FDTD applicability in lossy nanostructures becomes even more pertinent as researchers begin to build hybrid plasmonic/photonic systems~\cite{Benson2011}, which contain QDs and metal particles. 
Many colloidal QDs 
are also known to be described 
in terms of a finite-size dipole within a complex  index~\cite{UBC:2012:OE}.

In this Letter, we demonstrate that FDTD can be efficiently applied to compute the regularized GF in any inhomogeneous lossy medium.
We compare with known exact solutions and show excellent agreement. We also address the local-field problem, and again show that FDTD
can calculate the local-field ${\bf G}$  for any arbitrary structure.
As a representative example,
we consider a spherical metal nanoparticle (MNP) in air with a radius $a=20~$nm where the permittivity of the MNP is given via the Drude model, $\varepsilon = \varepsilon_r - \omega_p^2/(\omega^2 + i\gamma \omega)$, with  parameters  similar to silver: 
$\varepsilon_r=6$, $\omega_{p}/2\pi = 7.89$~eV, and $\gamma/2\pi = 51$~meV. We define a \emph{projected} local density of states (LDOS) along the $i$th axis as,
\begin{equation}
\label{eq:LDOS}
\rho_i(\br;\omega) = \frac{{\rm Im}[G_{ii} (\br,\br;\omega)]}{{\rm Im}[G_{ii}^{\rm vac} (\br,\br;\omega)]},
\end{equation}
where we have normalized by the  vacuum GF, so that this function directly gives  the Purcell factor~\cite{Purcell}.

\begin{figure}[!t]
\centering
\subfloat{
\includegraphics[width=0.9\columnwidth]{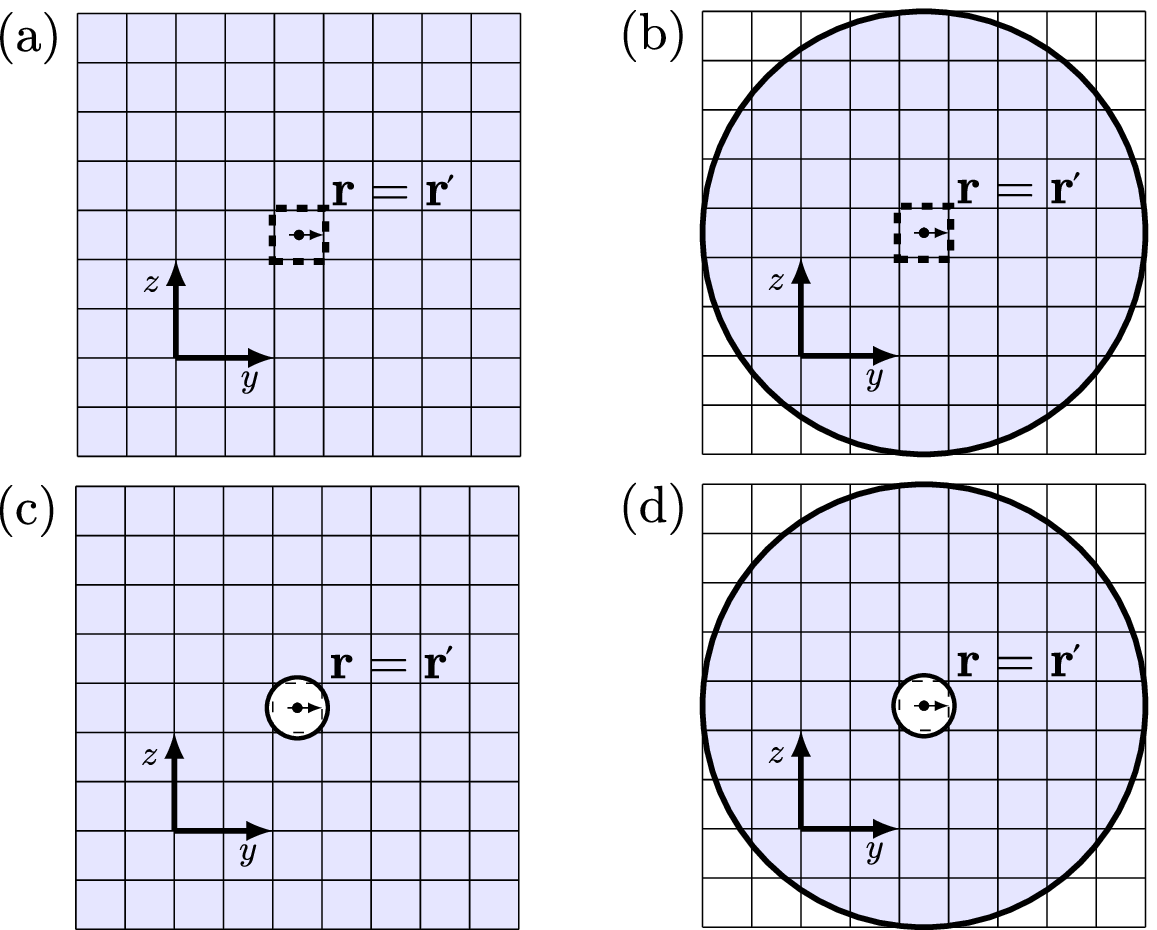}
}\\
\vspace{-0.2cm}
\subfloat{
\begin{overpic}[width=0.93\columnwidth]{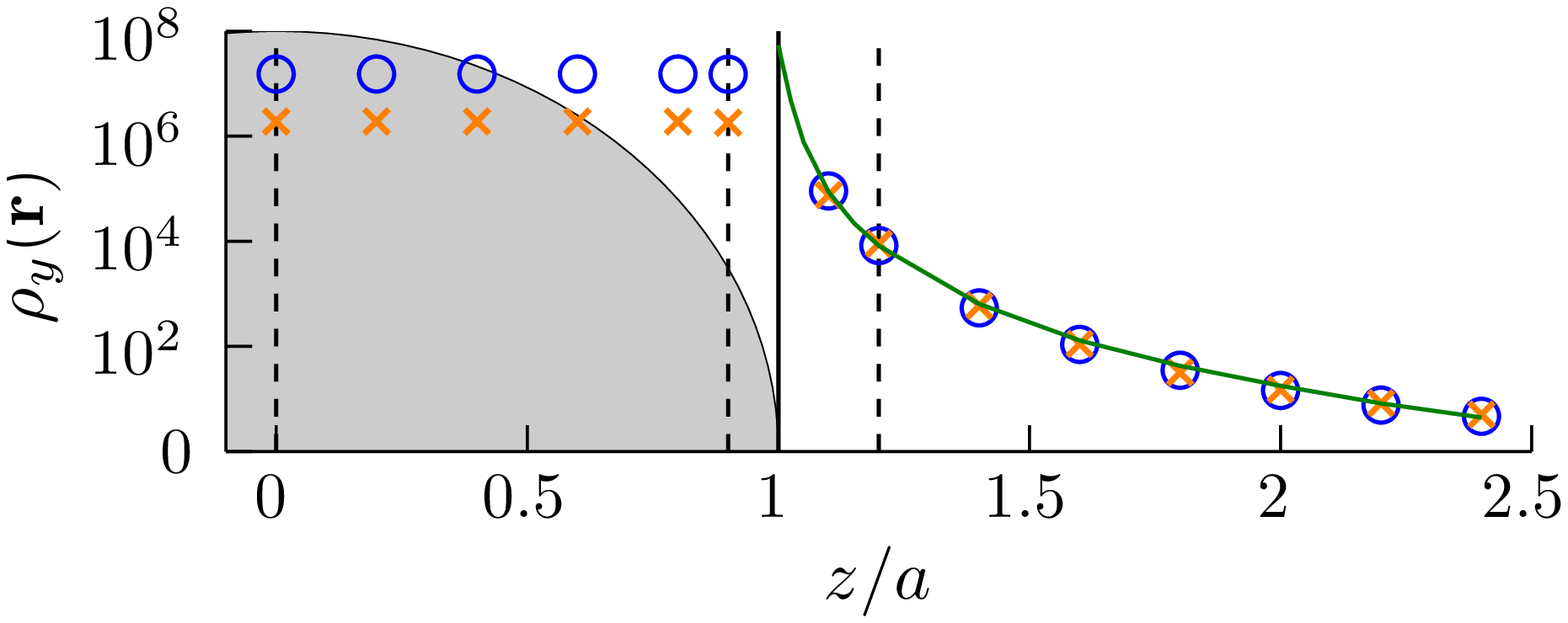}
\put(90,32){(e)}
\end{overpic}
}
\vspace{-0.3cm}
\caption{\label{fig:fig1} Schematics of the regularized GF integrated over a finite volume, for (a) a homogeneous material, and (b) a MNP. Schematics of the real cavity GF are shown inside (c) a homogeneous material, and (d) a MNP. (e) $\rho_{y} (\br)$ as a function of height above and inside a MNP. FDTD calculations (log scale) are show for 1-nm (blue circles) and 2-nm (orange crosses) grids. Outside  (above) the MNP, these are compared with the analytic GF (solid green line). The solid vertical black line shows the edge of the MNP, and the dashed vertical black lines show the cases considered in Fig.~\ref{fig:fig2}.}
\vspace{-0.6cm}
\end{figure}

In Fig.~\ref{fig:fig1} we show schematics of the geometries we will investigate. Figure~\ref{fig:fig1}(a) depicts a homogeneous space where we have divided the region onto a grid. For this scenario, we calculate the GF via two methods: ($i$) using FDTD and varying the FDTD grid size (Yee cell~\cite{Sullivan-FDTD}), and ($ii$) analytically integrating the homogeneous GF over a cubic integration volume~\cite{PhysRevE.70.036606}. Both of these methods act to \emph{regularize} the GF over a finite size.
The need to obtain regularized GFs is known to produce physically meaningful results, even in the context of point scatters
in free space~\cite{Lag}.
 Figure~\ref{fig:fig1}~(b) shows a similar calculation geometry as in (a) except we now consider a MNP, where the total GF can be broken into homogeneous parts [as in Fig.~\ref{fig:fig1}(a)] plus scattered parts~\cite{L.-W.Li1994}---and compared with the total GF as calculated using FDTD. Figure~\ref{fig:fig1}(c) shows the real cavity model where a small cavity of lossless dielectric (vacuum in this example) is placed inside a homogeneous space~\cite{Scheel-PRA-60-4094} which can be decomposed into homogeneous parts, Eq.~(\ref{eq:Ghom}), and scattering parts using analytical methods~\cite{L.-W.Li1994}; or the total GF can be calculated using FDTD. Figure~\ref{fig:fig1}~(d) shows the calculation geometry of the real cavity model inside a MNP of radius $a$, where the emitter cell has a different
(and real) refractive index; this latter geometry can be decomposed into a spherical multilayer problem~\cite{Tomas-PRA-63-053811}, or solved directly via FDTD. We stress that we can apply the FDTD to any arbitrary shaped
structure, but we choose a spherical geometry to compare with
known analytical solutions.

Figure~\ref{fig:fig1}(e) shows the peak LDOS along the $y$ direction,  $\rho_{y} (\br,\omega_{\rm peak})$, for the MNP as a function of height along the $z$ direction. We plot FDTD results using blue circles (orange crosses), corresponding to a gridding of $a/20$ ($a/10$), both of which corresponds to the scenario shown in Fig.~\ref{fig:fig1}(b). We use FDTD Solutions~\cite{Lumerical} and employ conformal meshing~\cite{883505} which is used to reduce staircasing effects due to the rectangular grid. We see that inside the MNP ($z/a<1$) the LDOS peak is almost constant up until the very edge of the MNP, indicating that the homogeneous contribution dominates this region. Additionally, the two grid sizes give different values for the LDOS {\em only inside} the MNP, despite ensuring we have more than 20 FDTD grid points-per-wavelength.
Different results for reduced grids typically imply  poor numerical convergence. Outside the MNP, we see that both FDTD grids give excellent agreement with each other and with the GF calculated using an analytical scattering technique~\cite{L.-W.Li1994}. This confirms FDTD computes the correct regularized GF for a finite-size dipole.

\begin{figure}[t]
\centering
\begin{overpic}[width=0.93\columnwidth]{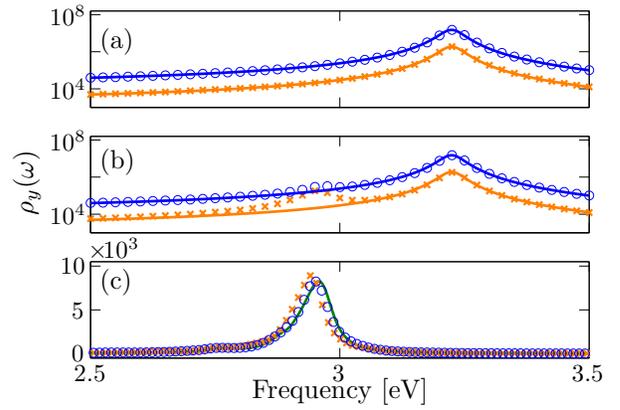}
\put(15,20){(c)}
\put(15,40){(b)}
\put(15,60){(a)}
\end{overpic}
\vspace{-0.2cm}
\caption{\label{fig:fig2} $\rho_{y} (\omega)$ as a function of frequency inside and outside the MNP. Blue circles (orange crosses) show a gridding of $a/20$ ($a/10$). The integration of the homogeneous GF over a cube of size $a/20$ ($a/10$) are given by blue-dark (orange-light) lines in (a-b). (a) $z/a=0$ in log scale. (b) $z/a=0.9$ in log scale. (c) $z/a=1.2$ in linear scale;  here, the green solid line shows   the analytic results. }
\vspace{-0.5cm}
\end{figure}

In Fig.~\ref{fig:fig2} we examine the LDOS as a function of frequency corresponding to the scenario shown in Fig.~\ref{fig:fig1}~(b), for three different heights, Fig.~\ref{fig:fig2}(a): $z/a=0$, Fig.~\ref{fig:fig2}(b): $z/a=0.9,$ and Fig.~\ref{fig:fig2}(c): $z/a=1.2$ [indicated by the dashed lines in Fig.~\ref{fig:fig1}(e)]. We plot FDTD results in each as blue circles for a gridding of $a/20,$ and orange crosses for a gridding of $a/10$. Additionally, we plot the regularized homogeneous GF calculated by integrating the homogeneous GF over a cubic volume of the exact same size as the FDTD grid, which we show as the solid lines in Fig.~\ref{fig:fig2}(a) and Fig.~\ref{fig:fig2}(b). We recognize that at the center of the MNP [Fig.~\ref{fig:fig2}(a)], the LDOS is dominated by the homogeneous contribution as there is no deviation between the homogeneous case and the total MNP case. In both cases the large peak at 3.23~eV corresponds to ${\rm Re}[\varepsilon]=0$. We also see that the formal integration of the GF agrees perfectly (to within less than a percent) for \emph{both} grids demonstrating again that the grid-dependent LDOS is actually the result of a finite sized dipole. As we approach the surface of the MNP, we see deviations from the homogeneous solutions [e.g., see region near 2.9~eV in Fig.~\ref{fig:fig2}(b)]; these deviations are due to the many non-dipolar surface plasmon modes that exist at the surface of the MNP~\cite{PhysRevB.85.075303} and are spectrally located in the same region as they are outside the MNP [Fig.~\ref{fig:fig2}~(c)]. In both Fig.~\ref{fig:fig2}~(a)-(b) we see that the peak of the LDOS is $\approx10^7$ the value of the vacuum GF (for $a/20$), which indicates that the homogeneous GF would give spontaneous emission enhancements (Purcell factors) which are orders of magnitude larger than any Purcell factors reported for semiconductor microcavities, though most of the emission is into non-radiating modes. 

\begin{figure}[t]
\centering
\vspace{-0.25cm}
\includegraphics[width=0.93\columnwidth]{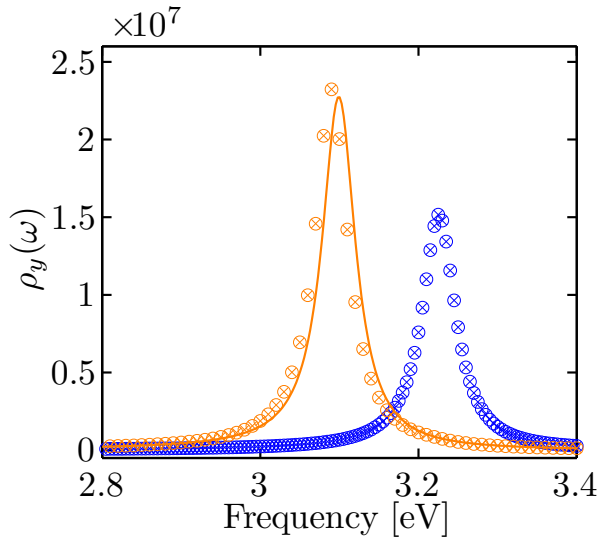}
\caption{\label{fig:fig3}  Local field calculations (real cavity) versus regularized GF calculations. In blue (right) we show the regularized
 GF for the homogeneous case [circles, Fig.~\ref{fig:fig1}(a)] and for the total GF inside a MNP [crosses, Fig.~\ref{fig:fig1}(b)] using FDTD with a gridding of $a/20$. In orange (left) we show the local field GF for the homogeneous case [circles, Fig.~\ref{fig:fig1}(c)] and for the total GF inside a MNP [crosses, Fig.~\ref{fig:fig1}(d)] using FDTD with a gridding of $a/20$, where we have inserted a spherical cavity filled with vacuum where the cavity volume gives the same effective volume as an FDTD grid cell, $(a/20)^3$. The orange solid line corresponds to exact calculations of the homogeneous local field GF using analytical techniques [Fig.~\ref{fig:fig1}(c)].}
\vspace{-0.4cm}
\end{figure}

When we examine the LDOS outside the MNP, we observe that the difference between the two FDTD grids are much reduced compared with inside, and they agree very well with analytical techniques (green solid line). The peak Purcell factor here is $<10^4$ which is three orders of magnitude smaller than inside the MNP for the $a/20$ grid. We also recognize that at such short distances, the  medium response is dominated by higher order plasmon modes, and the dipolar localized surface plasmon mode at 2.77~eV is again a factor of 10 smaller. The clear agreement between FDTD with analytic techniques is encouraging for the calculation of spontaneous emission rates in arbitrary lossy material geometries. This is especially true as we show FDTD is capable of calculating the GF both inside and outside a  spherical MNP, which is one of the most difficult systems for FDTD given the underlying rectangular grid and the large index contrast.

Next we examine local field effects to contrast with direct regularization of the GF, and to help describe the scenario of an embedded photon emitter with a different
refractive index (e.g., a QD in a metal MNP). In Fig.~\ref{fig:fig3} we compare FDTD results for each of the four scenarios as shown in Figs.~\ref{fig:fig1}~(a)-(d) using a grid size of $a/20$. For the local-field cavity, we insert a small region of vacuum at the center of the MNP where the radius of the cavity yields the same effective volume as an FDTD grid cell, namely $(a/20)^3$. The homogeneous GF (circles) and the total GF (crosses) are shown for both the regularized case (blue-right) and the local field case (orange-left). We see that the local field case exhibits an increase, as well as a frequency shift when compared with the regularized GF, which is the result of the fundamentally different scattering geometry. To ensure these discrepancies are not caused by FDTD numerics, we calculate the homogeneous local field case in Fig.~\ref{fig:fig1}(c) (solid orange line)  using exact analytical scattering techniques~\cite{Scheel-PRA-60-4094,L.-W.Li1994} and see very good agreement with the local field FDTD results. Interestingly, in the local field case, the local field homogeneous case is still the dominant contribution just as in the regularized GF case and is the same for spherical or cuboid (not shown) cavities. For cavities of real $\varepsilon_{\rm QD}=12$ (typical for semiconductor QDs) the real cavity resonance retains the same height, but shifts to 2.24~eV.

We summarize the regularization techniques as follows.
For the case where an emitter is located in a lossy material, but with no dielectric mismatch, then the regularized GF provides the proper method of regularization. However, for emitters with a dielectric mismatch such as QDs or atoms embedded in lossy host materials,  the local field GF provides the correct manner of regularization. For the case when the QD itself is lossy, and is embedded inside a lossy material, then \emph{both} regularization and local field effects must be considered.

In conclusion, our numerical calculations presented here have shown that FDTD is an excellent tool for the calculation of spontaneous emission rates in and near lossy materials, and that the technique correctly takes into account regularization and local field effects. It is also one of the few tools available which allows the calculation of the GF in completely arbitrary geometries. 


\bibliographystyle{osajnl}

\end{document}